TYPE Perspective
PUBLISHED 21 October 2022
DOI 10.3389/fphy.2022.1050277

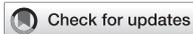

# Methods in econophysics: Estimating the probability density and volatility

Moawia Alghalith*

Economics Department, UWI, St Augustine, Trinidad and Tobago

We discuss and analyze some recent literature that introduced pioneering methods in econophysics. In doing so, we review recent methods of estimating the volatility, volatility of volatility, and probability densities. These methods will have useful applications in econophysics and finance.

## 1 Introduction

The volatility estimation is a key topic in finance and econophysics. It is an indicator of the movement in the asset price. For example, see [1,2]. Recently, the literature focused on the volatility of volatility. Examples include [3,4]. Closely related to the volatility estimation is the probability density estimation. The density estimation can be used to estimate the volatility and volatility of volatility. Needless to say, the probability density has many other applications.

In this note, we briefly discuss recent methods in the estimation of the volatility, volatility of volatility, and probability densities.

## 2 Review

There are typically two methods of density estimation: parametric and non-parametric methods. For example, [5,6] adopted the parametric method. [7–13] used the non-parametric approach. [14–17] provided empirical estimation. [18] used copulas. [19] used histograms and numerical simulations. [20] employed orthogonal polynomials.

A limitation of the parametric method is that it requires knowing the marginal distributions [5,21]. While the bandwidth selection problems, the high computational cost, and the kernel specification are some of the limitations of the non-parametric approach.

In response to some of these limitations, [22] introduced non-parametric methods for estimating the marginal and joint probability densities. The advantage of these methods is their relative simplicity. In particular, it allows us to circumvent the bandwidth selection problem and the kernel specification. Accordingly, the joint density can be calculated as

$$f(x, y) = \frac{\triangle^2 F(x, y) - \triangle f_X(x) \triangle x - \triangle f_Y(y) \triangle y}{2 \triangle x \triangle y}, \quad (1)$$

where $F(x, y)$ is the joint cumulative density, $f(x, y)$ is the joint density, $f_X(x)$ and $f_Y(y)$ are the marginal densities, $x$ is the outcome of $X$, $y$ is the outcome of $Y$, and $\triangle$ is the difference operator. The limitation of this method is that it requires high-frequency data for a high level of accuracy.

Using Taylor's expansions, [6, 23] introduced parametric methods for estimating the joint, marginal, conditional, and cumulative probability densities. In doing so, they relied on estimating regressions. For example, the joint density can be given by

$$f(x, y) = c_1 + c_2 x + c_3 y + c_4 x^2 + c_5 y^2 + c_6 x y, \quad (2)$$

where $c_i$ is a constant. The marginal density can be obtained by integrating the above equation.

The advantage of this method is its simplicity and the fact that the marginal distributions need not be known. Moreover, the estimation accuracy can be improved by increasing the order of the Taylor expansion. The limitation of this method is that we need to ensure the goodness-of-fit of the regression.

Previous literature on volatility typically considered time series, such as the generalized autoregressive conditional heteroskedasticity GARCH models. For example, see the excellent surveys by [1,2,24]. Asai and McAleer [25] adopted a Wishart stochastic volatility model. Bollerslev et al (2011) investigated asymmetry in volatility. Asai et al [26] assumed a noisy realized volatility. Muhle-Karbe et al [27] considered multivariate stochastic volatility. Sahiner [28] used the GARCH method. Mastroeni [29] considered vanishing stochastic volatility.

Alghalith [4] provided novel, parametric methods for estimating the volatility and volatility of volatility. According to this model, volatility data are needless. Also, the method can be applied to cross-sectional data. Furthermore, estimating the volatility matrix can be avoided. The limitation of this model is that we need to ensure the validity of the non-linear regression results.

Alghalith et al [30] introduced a simple, non-parametric method to estimate both the volatility and volatility of volatility. Accordingly, the volatility of the asset returns and volatility of volatility can be estimated, respectively, as

$$v_t = \sqrt{\frac{(\Delta S_t)^2}{S_t^2}}, \quad (3)$$

where $S_t$ is the price of the asset (typically a stock) at time $t$ and $v_t$ is the estimated volatility at time $t$.

$$\gamma_t = \sqrt{\frac{(\Delta v_t^2)^2}{v_t^2}}, \quad (4)$$

where $\gamma_t$ is the estimated volatility of volatility at time $t$.

Also, [30] explored the possibility that the volatility of volatility is not constant. The advantage of this approach is its simplicity. Its limitation is that it requires high-frequency data for a high level of accuracy.

In sum, these methods introduced by Alghalith and co-authors were reasonably accurate when they were applied to practical examples. In general, they were at least as accurate as other methods. However, the accuracy can be improved by increasing the frequency of the data or the order of the Taylor expansion.

## 3 Conclusion

We introduced simpler and less restrictive methods for estimating the volatility, volatility of volatility, and probability densities. In general, the other methods are more technical. Future research can utilize these methods of density estimation to estimate the volatility and volatility of volatility. Moreover, future research can apply these methods to other areas of econophysics.

## Data availability statement

The original contributions presented in the study are included in the article/supplementary material, further inquiries can be directed to the corresponding author.

## Author contributions

The author confirms being the sole contributor of this work and has approved it for publication.

## Acknowledgments

I'm very grateful to Editor SS and the reviewers for their excellent and fast comments.

## Conflict of interest

The author declares that the research was conducted in the absence of any commercial or financial relationships that could be construed as a potential conflict of interest.

## Publisher's note

All claims expressed in this article are solely those of the authors and do not necessarily represent those of their affiliated organizations, or those of the publisher, the editors and the reviewers. Any product that may be evaluated in this article, or claim that may be made by its manufacturer, is not guaranteed or endorsed by the publisher.